\documentclass[times,twocolumn,authoryear,final]{elsarticle}
\usepackage{jasr}
\usepackage{framed,multirow}

\usepackage{amssymb,amsmath,graphicx}
\usepackage{latexsym}

\usepackage[switch]{lineno}

\usepackage{url}
\usepackage{xcolor}
\definecolor{newcolor}{rgb}{.8,.349,.1}
\usepackage{colortbl}

\usepackage{siunitx} % For \unit and other SI unit formatting

\usepackage{booktabs}

\usepackage[citebordercolor=white]{hyperref}
\usepackage{amssymb}% http://ctan.org/pkg/amssymb
\usepackage{pifont}% http://ctan.org/pkg/pifont
\journal{Advances in Space Research}

\begin{document}

\verso{R.J.S Airey \textit{et. al.}}

\begin{frontmatter}

\title{Brightness, Colour, Polarisation: A Multi-Instrument Observation Campaign of the Gorizont-6 Satellite}%
    \author[Warwick-astro,Warwick-csda]{Robert J S \snm{Airey}}
    \ead{Robert.Airey@warwick.ac.uk}
    \address[Warwick-astro]{Department of Physics, University of Warwick, Coventry, CV4 7AL (UK)}
    \address[Warwick-csda]{Centre for Space Domain Awareness, University of Warwick, Coventry, CV4 7AL (UK)}
    \address[warwick-ceh]{Centre for Exoplanets and Habitability, University of Warwick, Gibbet Hill Road, Coventry CV4 7AL, UK}
    \author[Warwick-astro,Warwick-csda]{Paul \snm{Chote}}
    \author[herts]{Klaas \snm{Wiersema}}
    \address[herts]{Centre for Astrophysics Research, University of Hertfordshire, Hatfield, AL10 9AB (UK)}
    \author[Warwick-astro,Warwick-csda,warwick-ceh]{Ioannis \snm{Apergis}}
    \author[Warwick-astro,Warwick-csda,warwick-ceh]{James \snm{McCormac}}
    \author[Warwick-astro,Warwick-csda]{James A \snm{Blake}}
    \author[Warwick-astro,Warwick-csda]{Benjamin F \snm{Cooke}}
    \author[Warwick-astro,warwick-ceh]{Isobel S \snm{Lockley}}
    \author[Warwick-astro,warwick-ceh]{Peter J \snm{Wheatley}}
    \author[Warwick-astro,warwick-ceh]{Daniel \snm{Bayliss}}
    \author[Warwick-astro,warwick-ceh]{Samuel \snm{Gill}}
    \author[Belfast-astro]{Christopher A \snm{Watson}}
    \address[Belfast-astro]{Astrophysics Research Centre, School of Mathematics and Physics,
    Queen’s University Belfast, Belfast, BT7 1NN, UK}
    
    \author[brera,como]{Stefano \snm{Covino}}
       \address[brera]{INAF – Osservatorio Astronomico di Brera, Via E. Bianchi 46, 23807 Merate (LC), Italy}
       \address[como]{Como Lake centre for AstroPhysics (CLAP), DiSAT, Università dell’Insubria, via Valleggio 11, 22100 Como, Italy}
       \author[leiden]{Frans  \snm{Snik}}
       \address[leiden]{Leiden Observatory, Leiden University, P.O. Box 9513, 2300 RA Leiden, The Netherlands}
      \author[ljmu]{Jon  \snm{Marchant}}
          \address[ljmu]{Astrophysics Research Institute, Liverpool John Moores
University, IC2, Liverpool Science Park, 146 Brownlow Hill,
Liverpool, L3 5RF, UK}
           \author[rhul]{Justyn  \snm{Maund}}
\address[rhul]{Department of Physics, Royal Holloway - University of London, Egham, TW20 0EX, UK}
        \author[lancaster]{Brooke \snm{Simmons}}
        \address[lancaster]{Department of Physics, Lancaster University, Lancaster, LA1 4YB, UK}
        \author[ljmu]{ Iain A. \snm{Steele}}

\received{}
\finalform{}
\accepted{}
\availableonline{}

\begin{abstract}
We present a coordinated multi-instrument photometric and polarimetric study of the defunct geosynchronous satellite, Gorizont-6. This observation campaign combined wide-field multi-colour observations with simultaneous multi-site photometry and linear polarimetry. Our results demonstrate that the combined simultaneous colour and polarimetric measurements aid in breaking the degeneracy between the fundamental spin period and its harmonics, enabling light-curve features to be associated with specific reflecting surfaces. Using phase-dispersion minimisation with bootstrap resampling, we then measure a steadily increasing rotation period across six epochs spanning nine months, well described by a damped exponential curve consistent with Yarkovsky--O'Keefe--Radzievskii--Paddack (YORP) driven spin-down. A geometrical analysis of the multi-site observations provides additional constraints on the spin-axis orientation based on paired glint events. Overall, multi-colour photometry yields efficient, robust period measurements compared with single-band data, while polarimetry supplies the decisive constraints needed for unambiguous rotation-state determination, highlighting the value of combined photometric–polarimetric strategies for characterising defunct satellites.
\end{abstract}

\begin{keyword}
%% Keywords
\KWD Multi-colour photometry\sep Multi-site photometry\sep Polarisation\sep Period determination\sep Spin-state\sep Satellites
\end{keyword}

\end{frontmatter}

%% For linenumbers
%\linenumbers

\section{Introduction}
The population of defunct non-cooperative satellites in the near-Earth orbital environment continues to grow, affecting the regime as a sustainable space \citep[see e.g.,][]{2022A&G....63.2.14B, 2024AcAau.217..238B}. Defunct objects are likely to pose risks to the active population of satellites; therefore, it is important to characterise and understand their rotational dynamics. Orbiting objects that are not stabilised will begin to tumble under the influence of external torques. One such torque is the Yarkovsky--O'Keefe--Radzievskii--Paddack (YORP) effect, the physical basis of which originates from early radiation-pressure studies by \citet{1951PRIA...54..165O}, \citet{1952AZh....29..162R}, \citet{1969JGR....74.4379P}, and \citet{OKeefe1976}. The first modern formulation of the effect was given by \citet{2000Icar..148....2R}, demonstrating that asymmetric reflections and the delayed thermal re-emission of absorbed sunlight generate a net torque capable of altering an object's spin rate and obliquity. Subsequent theoretical developments extended YORP modelling to non-uniform, irregular, or tumbling bodies \citep[e.g.][]{2007Icar..188..430S,2010MNRAS.401.1933B,2010CeMDA.106..301C,2011MNRAS.417.2478B}. Observational detections of YORP-driven spin evolution for near-Earth asteroids---including (54509) YORP \citep{2007DPS....39.0502L} and Bennu \citep{2014Icar..235....5C,2019NatCo..10.1291H}---demonstrate that sunlight-driven torques can dominate long-term rotational dynamics. Given the large asymmetric geometries of many defunct satellites, similar YORP-driven behaviour is expected to contribute to their secular rotational evolution and possibly fragmentation due to extreme spin-up. This behaviour has been modelled dynamically as YORP-induced rotational fission for near earth asteroids \citep[e.g.][]{2011Icar..214..161J,2013DPS....4530108R}, but fragmentation as a result of the YORP effect has not been observed for resident space objects (RSOs). 

Utilising photometric measurements to estimate the spin state of RSOs has been an active area of research \citep[see e.g.][]{ 2018AdSpR..61.2135E,2019amos.confE..52C,2021AdSpR..67..360B,2024AdSpR..74.5725K,2024amos.conf..115M,2025RASTI...4af058A,2026AcAau.242...85T}. The rotational motion of these objects produce periodic signatures in their observed optical light curves, from which rotational periods can be derived. Multiple observations can reveal the state of dynamics for the object if, for example, the period and observed light curve significantly change \citep{PAPUSHEV20091416}. However, a key challenge remains when identifying the true rotation period. Harmonics (e.g.\,$P/2$, $2P$) of the fundamental period, $P$ can often yield similar phase-folded light curves that period-determination techniques such as Lomb--Scargle and phase-dispersion minimisation (PDM) can struggle to differentiate. Modelling the optical characteristics of the satellite can help to resolve these ambiguities in period \citep{2014amos.confE..61B}. 

The observed light curves can also reveal insights into the nature of the rotational dynamics taking place, whether that is uniform rotation about a given spin axis or rotation due to non-principal-axis (NPA) motion \citep[e.g.][]{2017amos.confE..62B}. Multi-site observations can also be used to infer the spin-axis of slow-spinning space objects \citep[e.g.][]{2020OAP....33..119K,Reichegger2022MultistaticLightCurves}. In parallel, recent multicolour and multi-colour photometric studies have shown the potential of such measurements for RSO characterization and classification \citep[see e.g.][]{2022AdSpR..70.3311Y,2024AdSpR..73.6161C,ZIGO20257365}. In this study, we investigate the combined impact of simultaneous multi-colour and polarimetric measurements toward breaking the degeneracy between the fundamental rotation period and its harmonics, and utilise long-baseline simultaneous two-site observations to probe possible spin-axis solutions.

\subsection{Gorizont-6 Background}

Gorizont-6 (NORAD: 13624, COSPAR: 1982-103A, also known as Gorizont No. 16L) is a decommissioned Soviet geostationary communications satellite launched in 1982 and raised into a graveyard orbit after its operations ended in 1989 \citep{Wade_Gorizont}.

\begin{figure}[ht!]
    \centering
    \includegraphics[width = 0.5\textwidth]{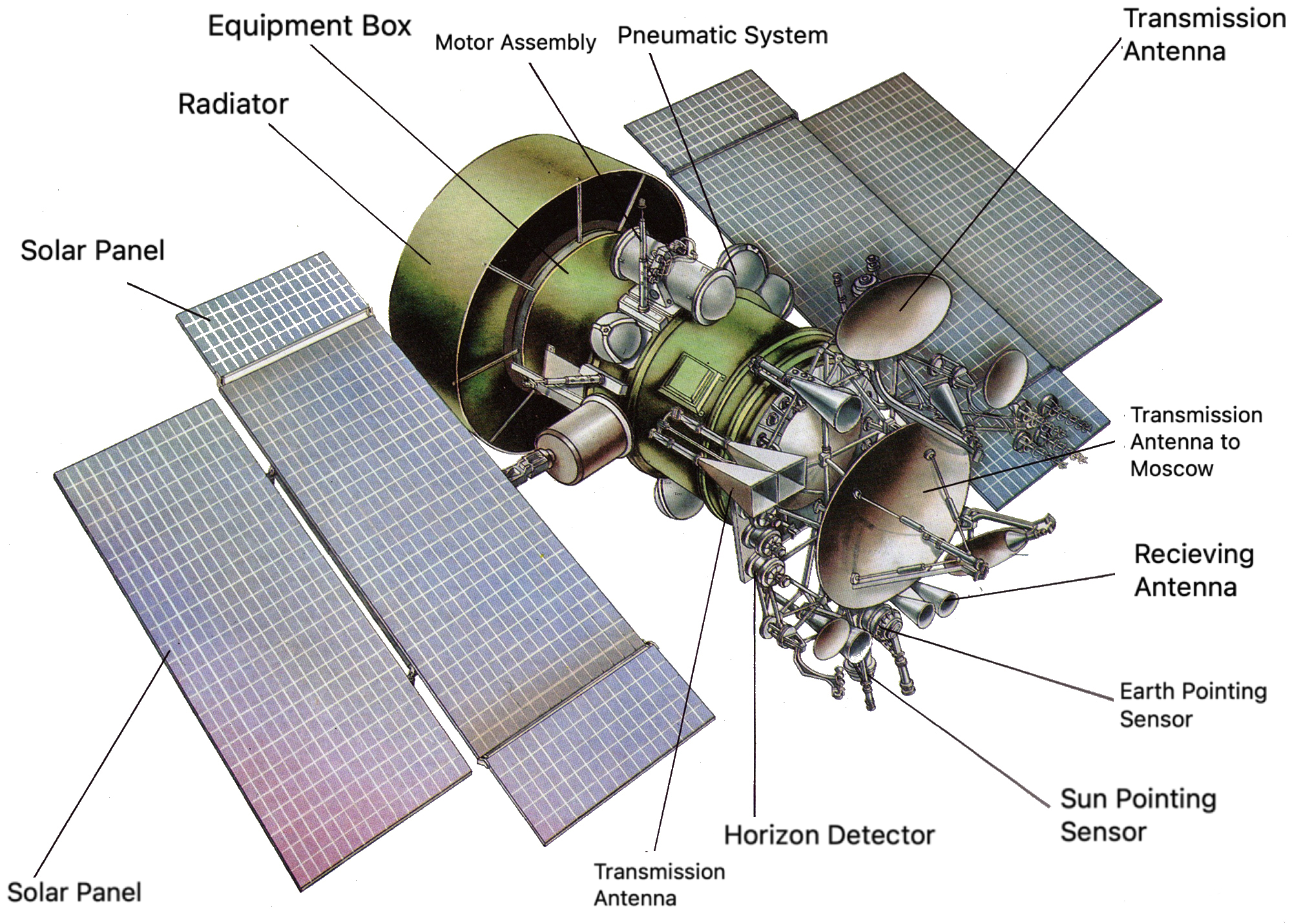}
    \caption{An example illustration of one of the satellites belonging to the Gorizont family. The key features of the KAUR-3 bus design are annotated. This figure is reproduced and translated from \cite{kosmonavtika_gorizont}.}
    \label{fig:Gorizont-6}
\end{figure}

The family of Gorizont satellites all have a common design built upon the KAUR-3 bus, shown in Fig. \ref{fig:Gorizont-6}. This design has a cylindrical body with radius of 2 m and length of 5 m and two solar panels that are approximately 3.73 m × 5.45 m each \citep{2015PhDT.......147A}.

Gorizont-6 was chosen for this study due to its favourable position in the sky from both Chile and La Palma on the night of 12th December 2023, and because previous observations \citep{2019amos.confE..52C} indicated that Gorizont light curves generally feature sharp specular glints, which could potentially be used to measure parallax-induced timing offsets. Subsequently, the object was incorporated into a broader polarimetric survey of GEO graveyard objects (Wiersema et al., in prep.)

\subsection{Polarimetry}
For satellites and space debris, optical photometry and spectroscopy form the conventional basis for characterisation, constraining their attitude, surface properties and behaviour \citep[e.g.][]{2013amos.confE..34H,2015amos.confE..74J,2020AdSpR..65..326S,2023amos.conf...44M,2025AdSpR..76..764C}

In addition to multi-colour light curves and spectra, a powerful observational diagnostic to study objects in space is the optical linear polarisation of the received light. Reflection of (sun)light off a solid body induces linear polarisation because the Fresnel reflection coefficients differ for the field components parallel and perpendicular to the plane of incidence ($R_p \neq R_s$).
The main parameters that determine the amount of observed linear polarisation for specular reflection off a body in space are the refractive indices of the reflecting material(s), the reflection geometry (i.e. the angle(s) of the reflecting component(s) with respect to the observer and the Sun) and the shape (at both small and large size scales) of the reflecting surface. Additionally, the polarisation state of the light prior to reflection has an important influence; note that sunlight is generally unpolarised (< $10^{-7}$) \citep{1987Natur.326..270K}. This leads to the possibility of using optical linear polarimetry (as a function of solar phase angle) to study both the orientation of the reflector and its (surface) material properties.

In stabilised, active satellites, the polarisation properties have received significant observational attention \citep[see e.g.][]{Speicher2015_IdentificationGEOUsingPolarization,2016amos.confE..23C,WiersemaThor6,2022AdSpR..69..581P,2024JAnSc..71...39E}, but its true potential for debris objects is still mostly unexplored. As an object rotates, the angle(s) of the reflecting part(s) with respect to the Sun and observer rapidly change. Where light curves and colours can show degeneracy between different orientations of the object (particularly when the shape of the object is symmetric along one or more planes), the polarisation fraction and polarisation angle are not degenerate in the same way. Polarimetry may therefore provide a means to break spin state degeneracies present in light curves. If more information is known about the satellite itself and/or its spin state, polarisation may be able to uniquely identify the material and orientation of the reflecting components that contribute to specific light curve features. 

\section{Instrumentation and Observational Campaign}
\label{sec:instruments}
\begin{figure*}
    \centering
    \includegraphics[width=\textwidth]{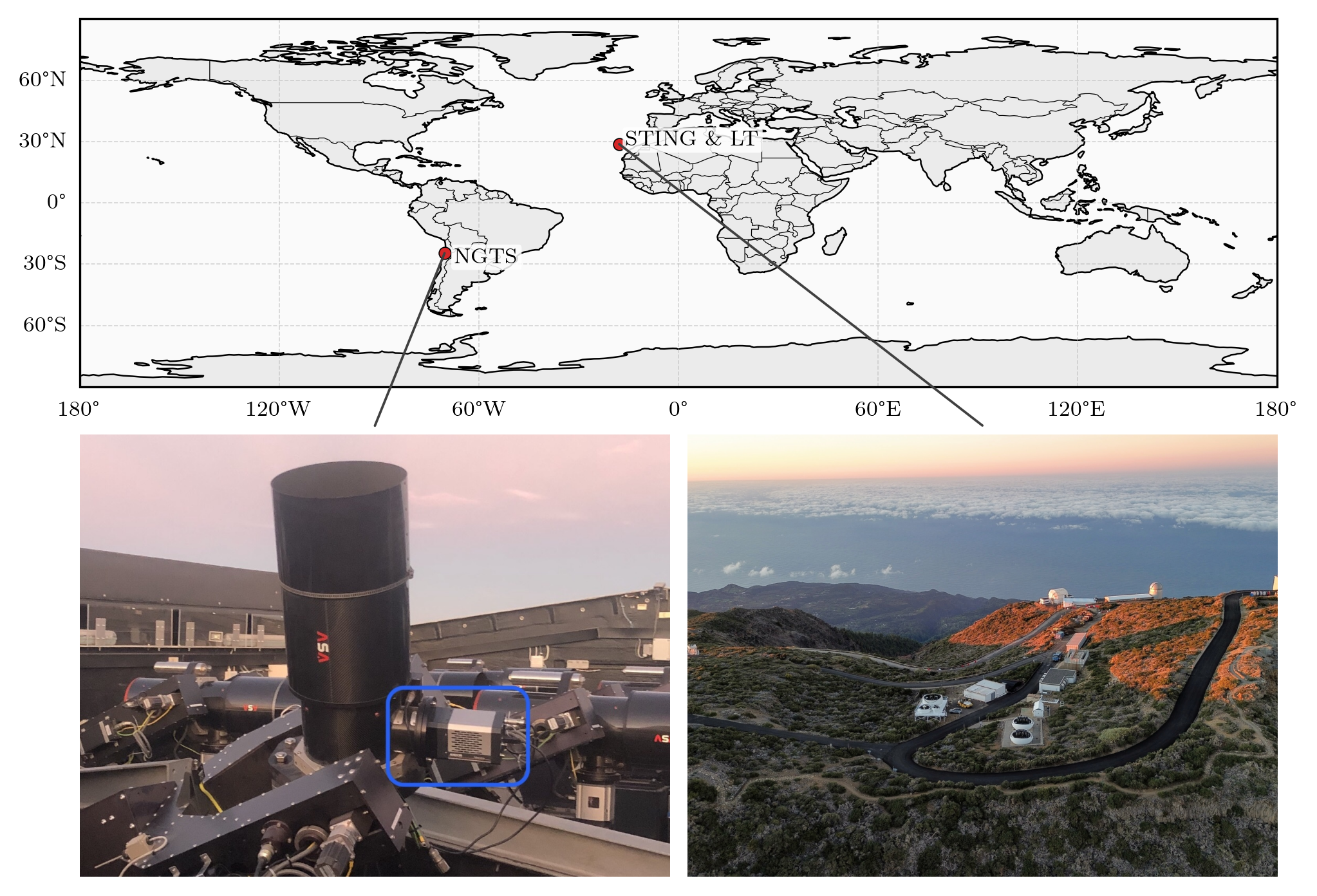}    \caption{\textbf{Top:} Cartographic map illustrating the site locations for the NGTS (Paranal), STING (La Palma) and LT (La Palma) instruments.  \textbf{Bottom Left:} A photograph of the Marana sCMOS camera, highlighted within the blue rectangle, mounted on one of the NGTS telescopes. This photograph is reproduced from \cite{2026arXiv260316361A} with permission. \textbf{Bottom Right:} A photograph of the ORM on La Palma, where STING and the LT are located.}
    \label{fig:sites}
\end{figure*}
Our observational campaign was designed around a baseline of simultaneous four-colour photometry, enhanced by two complementary techniques to strengthen rotation-state identification: multi-site high-cadence photometry and broadband linear polarimetry. On the 12th December 2023, this baseline was augmented by a simultaneous observation with NGTS \citep{10.1093/mnras/stx2836}, which provided long-baseline, second-site photometry and by the Liverpool Telescope (LT; \citealt{SteeleLT}) polarimetry, with the full combination of instruments used over several epochs shown in Table \ref{tab:my-table}.

STING \citep{2025AdSpR..75.5757A} and the Liverpool Telescope are located at the same observatory on La Palma, separated by less than 250 m. This ensures a virtually identical viewing angle geometry for satellites in or near the geostationary belt, allowing a direct comparison of observed features. Simultaneous observations from La Palma and Paranal (where the NGTS telescope is situated) provide a large baseline ($\sim$8170 km). This offers a different viewing geometry such that reflections from the spacecraft will be received with a noticeable phase difference ($\gg$ timing uncertainties). This can then be leveraged in order to make constraints on the spin-state of the satellite. The observing site locations for our observation campaign are illustrated in Figure \ref{fig:sites}.

\begin{table*}[h!]
\centering
\small
\begin{tabular}{|c|c|c|c|}
\hline
\textbf{Night Starting} & \textbf{NGTS [UTC]} & \textbf{STING [UTC]} & \textbf{LT [UTC]} \\
\hline

2023-12-12 &
00:34 -- 00:54 &
23:44 -- 02:29 &
00:12 -- 00:33 \\
\hline

2023-12-29 &
-- &
22:58 -- 01:59 &
22:25 -- 22:43 \\
\hline

2024-01-21 &
-- &
03:02 -- 04:27 &
04:03 -- 04:23 \\
\hline

2024-05-13 &
-- &
20:54 -- 23:59 &
22:50 -- 23:10 \\
\hline

2024-07-19 &
-- & -- &
21:34 -- 21:54 \\
\hline

2024-09-02 &
-- &
20:49 -- 22:21 &
-- \\
\hline

\end{tabular}
\caption{Overview of the Gorizont-6 observations. The UTC start and end times for each observation are provided for the La Palma night (i.e. times after midnight correspond to the following day).}
\label{tab:my-table}
\end{table*}

\subsection{STING}\label{sec:STING_data}
STING is a wide-field ($10^\circ\times7.5^\circ$) camera system located at the Roque de los Muchachos observatory (ORM) in La Palma, Spain. STING consists of four co-mounted lenses that observe simultaneously in four broad-band colours ($B_{RGB}$, $G_{RGB}$, $R_{RGB}$ and $i$). The raw STING images are processed using an automated pipeline \citep{2025AdSpR..76..764C} that applies standard image corrections (bias, dark, flat field), performs astrometric and photometric calibration on each frame, identifies targets based on an input TLE (Two Line Element) catalogue, and extracts photometric measurements plus associated uncertainties. Images were acquired with an exposure time of 10 seconds, with the four cameras being synced to within a few tens of milliseconds.

\subsection{NGTS}\label{sec:NGTS_data}
The Next Generation Transit Survey is a ground-based observatory located at ESO’s (European Southern Observatory) Paranal Observatory in Chile. The observatory consists of twelve 20-cm wide field Newtonian telescopes, focused on the discovery and characterisation of extra-solar planets. In December 2023, one of the telescopes was reconfigured with an Andor Marana CMOS camera as part of a study into the use of scientific CMOS devices for high-precision photometry \citep{2025arXiv251014484A, 2026arXiv260316361A}. The telescope setup hosting the Marana camera operated independently of the main NGTS system, controlled by the same \texttt{rockit} software that operates STING (and other Warwick facilities on the ORM, \citep[see e.g.][]{2023AdSpR..72..907C,2025MNRAS.tmp.1986M}). A modified version of the STING pipeline was used to reduce the images and extract single-colour photometry in the standard NGTS filter ($520-890\,$nm).

The observation of Gorizont-6 presented here was acquired during the on-site commissioning of this test instrument. The target was observed for a duration of $\sim20$ minutes with the High Dynamic Range mode and 5-second exposures on the night of 12$^{\mathrm{th}}$ December 2023. 

\subsection{Liverpool Telescope (MOPTOP)}
Polarimetric observations were obtained using the 2.0\,m Liverpool Telescope under proposal numbers PL23B03 and PL24A03 (PI Wiersema). We used the dual-beam imaging polarimeter MOPTOP (Multi-colour OPTimised Optical Polarimeter; \citealt{Jermak2016,Jermak2018,ShresthaMOPTOP}), following the observational strategy and data reduction described in \cite{WiersemaThor6}. Observations were acquired in {\tt SLOW} mode, giving a polarimetric cadence of 20\,s: in {\tt SLOW} mode the half-wave plate does a full rotation every 80 seconds, in which 16 exposures are obtained at $22.5^{\circ}$ steps, we use four half-wave plate angles for each polarimetric datapoint (\citealt{WiersemaThor6}). The  choice of {\tt SLOW} mode was driven primarily by operational constraints on the LT: {\tt FAST}-mode sequences are limited to a maximum duration of $\sim$800\,s (because of file transfer and storage constraints), whereas {\tt SLOW} mode allows uninterrupted time series of essentially arbitrary length. Since the object's rotational modulation period was not known \emph{a priori}, longer continuous sequences reduce the risk of sampling an incomplete cycle. In addition, the exposure times per angle in {\tt SLOW} mode are $\sim$10$\times$ longer than in {\tt FAST}, improving the S/N for the target outside of glints; {\tt FAST} is therefore only advantageous for very bright sources and/or signals on timescales much shorter than 800\,s.

The {\tt MOP-R} filter was chosen due to the high quality of the existing calibration values for instrumental polarisation and de-polarisation; the availability of extensive comparison data for active and inactive geostationary satellites (Wiersema et al.\ in prep.), its lower sensitivity to moonlight compared to bluer filters, and the instrument's superior sensitivity at these wavelengths.

We use the tabulated values on the MOPTOP website\footnote{\label{note1}https://telescope.livjm.ac.uk/TelInst/Inst/MOPTOP/} for the {\em cam3} and {\em cam4} instrument configuration. We constructed pseudo-light curves from the MOPTOP data by computing the average flux $f'_i = (f_{{\rm cam3},i} + f_{{\rm cam4},i} + f_{{\rm cam3},i+1} + f_{{\rm cam4},i+1})/4$ (in contrast to the flux \emph{differences} used to construct the polarimetric measurements; see \citealt{WiersemaThor6} for further details), where we use sets of two waveplate angles for each average flux point (i.e. waveplate angles $0^{\circ},22.5^{\circ}$ make one average flux point, $45^{\circ},67.5^{\circ}$ make the next, etc). This results in a light curve with twice the cadence of the polarimetric data, since beam-switching required for accurate polarimetric calibration is not needed when measuring Stokes $I$ (total intensity). We then compute uncalibrated instrumental magnitudes as $m = -2.5\log(f')$ and use an arbitrary zero point offset to align it to the STING photometry.

\section{Results and Analysis}

\subsection{Period Analysis}
\label{sec:period_analysis}
We investigated the rotational period of Gorizont-6 using a Phase Dispersion Minimisation (PDM) \citep{1978ApJ...224..953S} analysis. PDM is well suited to detecting periodic signals with non-sinusoidal light curve morphology \citep{2005Icar..175..194F} and therefore provides a robust alternative to standard Lomb--Scargle techniques.

For our analysis we implemented a custom PDM algorithm. An initial grid of candidate periods was defined based on visual inspection of each light curve, spanning 100--2400\,s and sampled linearly with 30{,}000 trial periods. The best-fit period was identified as the period that minimised the PDM statistic:

\begin{enumerate}
    \item For a given trial period \(P\), phase-fold the light curve and fit a linear baseline whose subtraction minimizes the PDM statistic \(\Theta(P)\).
    \item Assign the de-trended, phase-folded residuals to overlapping phase bins with approximately equal population.
    \item Compute the intra-bin variances and form the PDM statistic \(\Theta(P)\) from their ratio to the total variance.
    \item Repeat for all trial periods and select \(P_{\mathrm{best}} = \arg\min \Theta(P)\).
\end{enumerate}

We use conditional de-trending at each trial period because the baseline trend is small relative to the periodic signal and therefore cannot be robustly fit independently. While this pragmatic approach is common, it is not fully statistically rigorous: separating de-trending from period inference can mildly underestimate parameter uncertainties because baseline-model uncertainty is not propagated. A fully self-consistent treatment would fit a joint Bayesian model and marginalize over baseline parameters; we do not pursue that here.

Uncertainties on the derived period were estimated using bootstrap resampling. For each bootstrap realization, the light curve was resampled with replacement and the full PDM period search repeated. The resulting distribution of best-fit periods was approximately Gaussian; we therefore report the mean period and its standard deviation as the \(1\sigma\) uncertainty.

Further details of the PDM methodology and statistical foundations can be found in \cite{1997ApJ...489..941S}.

The periods recovered from the individual photometric bands and from the colour light curves are summarised in Table \ref{tab:periods_bootstrap}. In several observing epochs, particularly where the single-band light curves exhibit strong harmonic structure, the periods inferred from individual bands differ by factors of approximately two or four, indicating ambiguity between the fundamental rotation period and its harmonics. 

\begin{figure*}[ht!]
    \centering
    \includegraphics[]{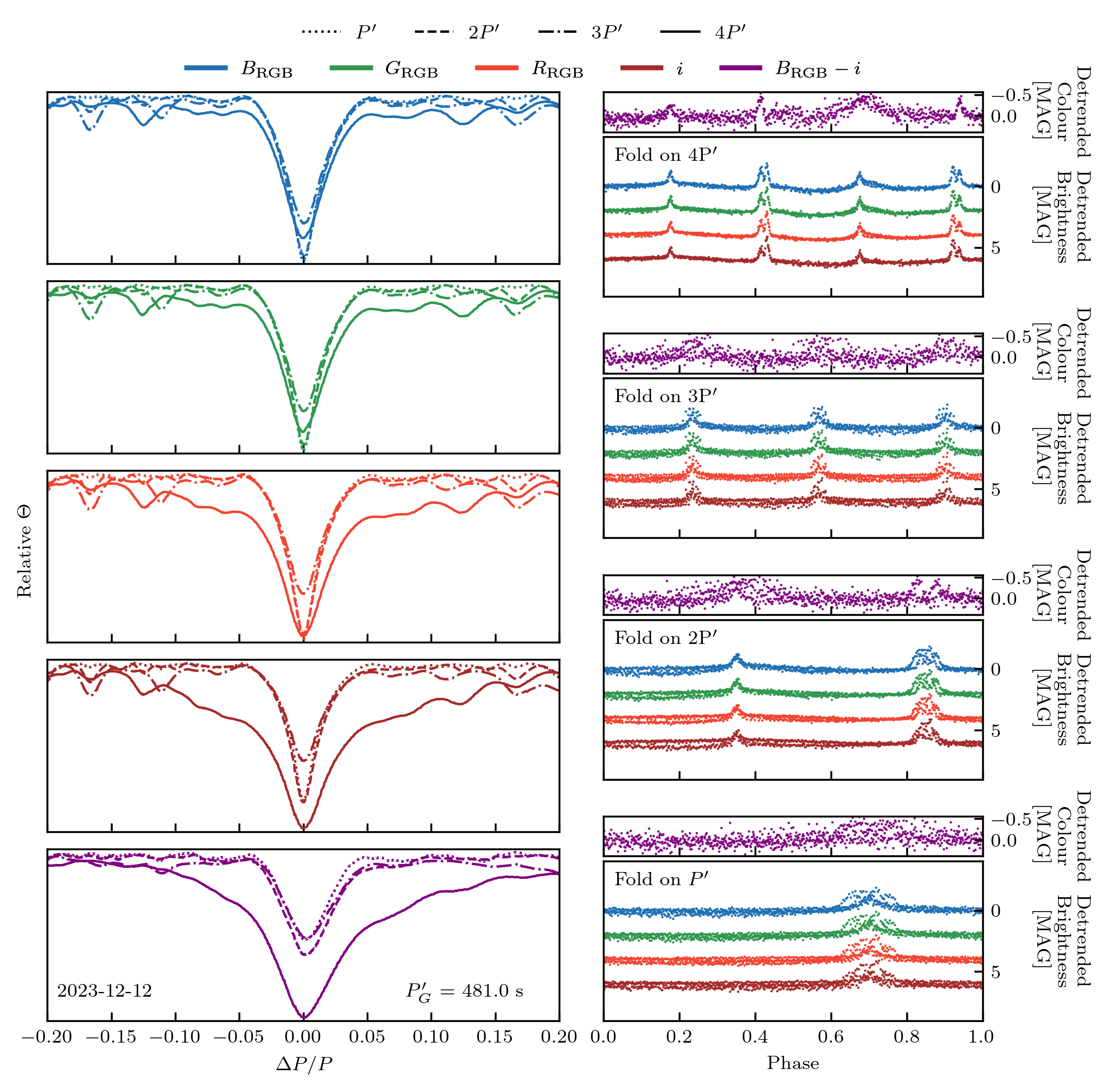}
    \caption{\textbf{Left:} Period--search statistic $\Theta(P)$ for the night 2023--12--12, shown as five panels (one per band: $B_{\mathrm{RGB}}$, $G_{\mathrm{RGB}}$, $R_{\mathrm{RGB}}$, $i$, and $B_{\mathrm{RGB}}-i$). In each panel we plot slices of $\Theta(P)$ centred on the \emph{trial} PDM period from the $G_{\mathrm{RGB}}$ band, $P'_G = 481.0\,\mathrm{s}$, and its first three harmonics ($2P'_G$, $3P'_G$, $4P'_G$) to demonstrate how well the PDM is able to distinguish between the different fold periods.
    \textbf{Right:} Phase-folded de-trended light curves in the $B$, $G$, $R$, and $i$ bands (with $B_{\mathrm{RGB}}-i$ shown above each panel), folded on the same set of trial periods $hP'_G$. Light curves are median-centred and vertically offset between bands for clarity; no additional amplitude scaling is applied.
}
    \label{fig:folded_stacked}
\end{figure*}

Fig. \ref{fig:folded_stacked} demonstrates this ambiguity between harmonics. We define the trial period
$P' \equiv P'_{G}$ as the PDM-derived period from the $G_{\mathrm{RGB}}$ band (here $P'_{G}=481.0\,\mathrm{s}$) (see Table \ref{tab:my-table});
as shown by the right-hand panels, folding on $P'$ yields a poor fold with clear bifurcations, indicating that
$P'$ is not the true rotational period but a harmonic of it. The choice of $P' \equiv P'_{G}$ is purely arbitrary as each band will share a similar trial period value for this fundamental harmonic.

The left column shows stacked, normalised PDM $\Theta$ curves centred on the first four harmonics
($P'$, $2P'$, $3P'$, $4P'$) for each band. Higher-order harmonics (>$\,4P'$) are increasingly disfavoured because folding over long trial periods approaches (or exceeds) the effective time baseline per segment, leading to poorly constrained phase coverage and spurious 
$\Theta$ structure; consequently, we focus on the first four harmonics.

In the single-band filters ($B_{\mathrm{RGB}}$, $G_{\mathrm{RGB}}$, $R_{\mathrm{RGB}}$, $i$)
there is comparatively little discrimination between these candidate periods (the $\Theta$ minima are broadly similar
across the four harmonics), whereas the colour curve $B_{\mathrm{RGB}}-i$ shows a much clearer preference for the
longer-period solution. We adopt the \(B_{\rm RGB}-i\) colour index (as opposed to other colour indices) as it spans the largest wavelength separation available in our
STING data, providing the strongest colour contrast between spectrally
distinct reflecting components (maximising the visibility of `bluer'/`redder'
excursions associated with different facets).

The right panels show the corresponding folded, de-trended light curves for each trial period $hP'$.
While the single-band folds remain double-valued or bifurcated at $P'$ (and are not fully resolved at $2P'$ and $3P'$),
folding on $4P'$ produces a coherent modulation across all bands and a consistent colour behaviour in
$B_{\mathrm{RGB}}-i$. We therefore adopt $P = 4P' \simeq 1926\,\mathrm{s}$ as the true rotational period for this epoch (21st January 2024).

The estimated period for the night in the $B_{\mathrm{RGB}}-i$ band is clearly unreliable due to the length
of the observation arc ($\sim5000\,\mathrm{s}$, the shortest of our epochs) being comparable to twice the estimated period;
we therefore manually select twice the $i$-band bootstrapped estimate for that night.

Overall, these results demonstrate that incorporating colour information into the period analysis can provide a more robust means of discriminating between the fundamental rotation period and its harmonics, particularly in cases where the single-band light curves alone yield ambiguous or multi-modal period solutions.

\begin{table}[ht!]
\centering
\begin{tabular}{llrr}
\toprule
Night & Band & $P_{\mathrm{best}}$ [s] & $P_{\mathrm{mean}} \pm \sigma_P$ [s] \\
\midrule
\multirow{5}{*}{2023-12-12}
 & $B_{\mathrm{RGB}}$          &  481.0  &  481.1 $\pm$ 0.43 \\
 & $G_{\mathrm{RGB}}$          &  481.0  &  481.1 $\pm$ 0.41 \\
 & $R_{\mathrm{RGB}}$          &  481.1  &  481.1 $\pm$ 0.47 \\
 & $i$                         & 1922.4  & 1924.5 $\pm$ 2.43 \\
 & $B_{\mathrm{RGB}} - i$      & 1926.0  & 1925.2 $\pm$ 3.12 \\
\midrule
\multirow{5}{*}{2023-12-29}
 & $B_{\mathrm{RGB}}$          &  501.6  &  501.3 $\pm$ 0.43 \\
 & $G_{\mathrm{RGB}}$          &  501.4  &  501.2 $\pm$ 0.45 \\
 & $R_{\mathrm{RGB}}$          &  501.7  &  501.2 $\pm$ 0.39 \\
 & $i$                         &  501.5  &  501.4 $\pm$ 0.48 \\
 & $B_{\mathrm{RGB}} - i$      & 2003.2 &  2001.2 $\pm$ 7.83 \\
\midrule
\multirow{5}{*}{2024-01-21}
 & $B_{\mathrm{RGB}}$          & 1022.9  & 1022.4 $\pm$ 3.12 \\
 & $G_{\mathrm{RGB}}$          & 1022.5  & 1023.5 $\pm$ 2.86 \\
 & $R_{\mathrm{RGB}}$          & 1025.2  & 1023.8 $\pm$ 2.52 \\
 & $i$                         & 1025.2  & 1023.1 $\pm$ 2.15 \\
 & \textcolor{blue}{$B_{\mathrm{RGB}} - i$}    & \textcolor{blue}{2400.0}  & \textcolor{blue}{2234.3 $\pm$ 162.27} \\
\midrule
\multirow{5}{*}{2024-05-13}
 & $B_{\mathrm{RGB}}$          & 1109.9  & 1109.2 $\pm$ 0.94 \\
 & $G_{\mathrm{RGB}}$          & 1109.5  & 1109.3 $\pm$ 1.20 \\
 & $R_{\mathrm{RGB}}$          & 1109.9  & 1111.1 $\pm$ 0.87 \\
 & $i$                         & 1109.9  & 1110.1 $\pm$ 0.76 \\
 & $B_{\mathrm{RGB}} - i$      & 2222.8  & 2223.4 $\pm$ 14.36 \\
\midrule
\multirow{5}{*}{2024-09-02}
 & $B_{\mathrm{RGB}}$          & 1127.9  & 1127.1 $\pm$ 2.73 \\
 & $G_{\mathrm{RGB}}$          & 1126.3  & 1127.8 $\pm$ 1.84 \\
 & $R_{\mathrm{RGB}}$          & 1125.9  & 1126.3 $\pm$ 2.17 \\
 & $i$                         & 1127.9  & 1126.1 $\pm$ 2.56 \\
 & $B_{\mathrm{RGB}} - i$      & 2254.1  & 2258.9 $\pm$ 9.73 \\
\bottomrule
\end{tabular}
\caption{PDM-derived periods and bootstrap mean periods with $1\sigma$ uncertainties for all observing STING observations and bandpasses (and $B_{\mathrm{RGB}}-i$ colour). The PDM arrived at the wrong harmonic solution (value highlighted in blue) for the night of the 21st of January 2024 in the $B_{\mathrm{RGB}}-i$ colour band. We therefore correct for this by manually selecting twice the $i$ -band bootstrapped estimate.}
\label{tab:periods_bootstrap}
\end{table}

\begin{figure}
    \centering
    \includegraphics[width=\linewidth]{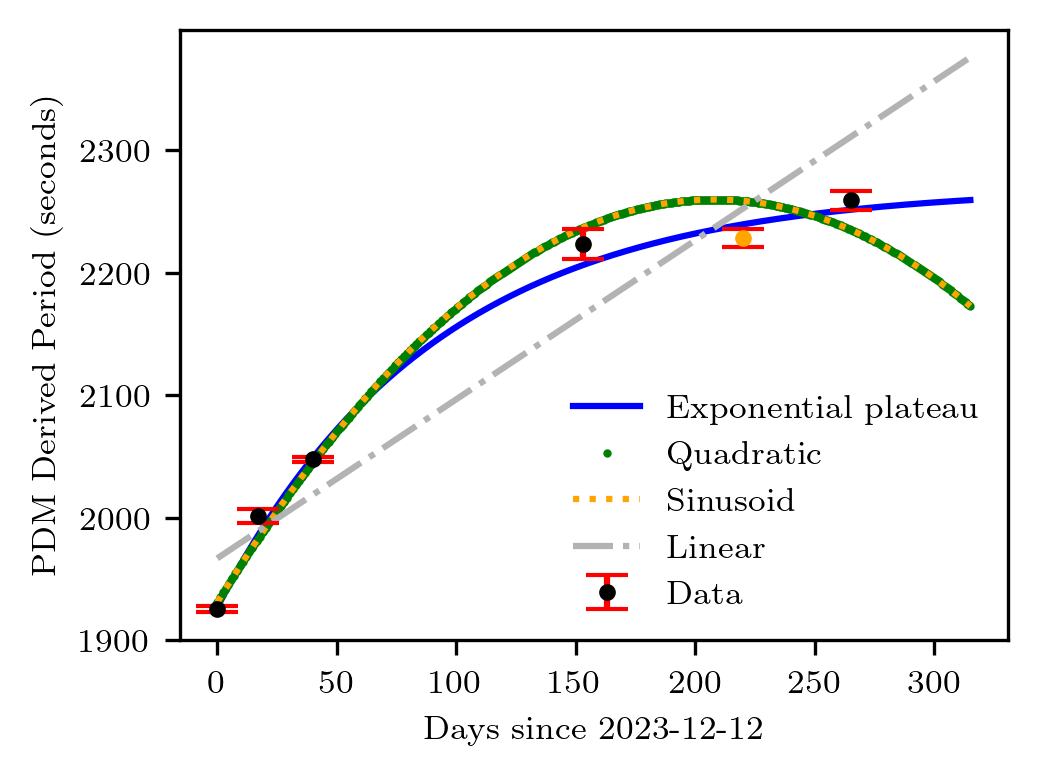}
    \caption{Evolution of the PDM-derived rotation period of Gorizont-6 as a function of time (days since the night of the 12th December 2023). Points show the best-fit periods from each observing epoch with 1$\sigma$ uncertainties; the solid curve shows the best-fitting exponential approach model (Equation \ref{eq:exp}). The MOPTOP observation is shown by the orange scatter point.}
    \label{fig:period_evolve}
\end{figure}

Fig.~\ref{fig:period_evolve} shows the evolution of the PDM-derived rotation period of Gorizont-6 as a function of time (days since the night of 12th December 2023), with each point representing the best-fit period from an observing epoch and vertical bars indicating the $1\sigma$ uncertainties. Over our observing baseline, the measured period increases asymptotically, suggesting a gradual approach toward a longer, quasi-stable spin period. One epoch (2024-07-19; orange scatter point) consists only of LT MOPTOP data; we estimated the period on the night via bootstrap resampling, restricting the period search bounds to lie between the lower (upper) $1\sigma$ limits of the preceding (following) observing epochs. In addition to the data, Fig.~\ref{fig:period_evolve} overlays several candidate trend models evaluated from weighted least-squares fits.

The asymptotic increase motivates an exponential approach model, which naturally captures a rapid early change followed by flattening toward a limiting value. This is supported by the goodness-of-fit: the exponential approach provides the smallest residual scatter among the functional forms tested, with $\chi^2 = 14.0$ for ${\rm DOF}=6-3=3$ (i.e.\ $\chi^2/{\rm DOF}=4.65$). By comparison, the alternative models in Fig.~\ref{fig:period_evolve} are substantially less consistent with the data given the quoted uncertainties (linear: $\chi^2/{\rm DOF}\approx 131$; quadratic: $\chi^2/{\rm DOF}\approx 15$; sinusoid: $\chi^2/{\rm DOF}\approx 23$). While $\chi^2/{\rm DOF}>1$ indicates that the scatter about even the best-fitting curve exceeds that expected from the formal $1\sigma$ errors---suggesting additional epoch-to-epoch variance and/or un-modelled systematics---the exponential form is nonetheless the most adequate empirical representation of the behaviour in this dataset.

Motivated by this, we quantify the long-term trend by fitting the data with an exponential approach model of the form
\begin{equation}
    f(x) = a - be^{-kx},
    \label{eq:exp}
\end{equation}
where $x$ is time in days since 12th December 2023 and $f(x)$ is the predicted spin period. The best-fitting exponential curve is shown as the solid line in Fig.~\ref{fig:period_evolve}. The fitted parameters are
\[
a = 2267.7 \pm 7.3 \,{\rm s}, \quad
b = 341.1 \pm 7.2 \,{\rm s}, \quad
k = 0.0110 \pm 0.0005 \,{\rm day}^{-1},
\]
with uncertainties corresponding to $1\sigma$ errors from the square roots of the diagonal elements of the fit covariance matrix.

In this parametrisation, $a$ is the asymptotic spin period approached within the current spin-cycle, $b$ sets the initial offset from this plateau, and $k$ controls the rate of approach. The associated e-folding timescale is $\tau = 1/k = 90.9 \pm 4.1\,{\rm days}$, meaning the difference $a - f(x)$ decreases by a factor of $e$ approximately every $\sim 91$ days. With the best-fit parameters, the implied initial period at the start of the baseline is $P(t{=}0)=a-b \approx 1927\,{\rm s}$, increasing toward an asymptotic value of $\approx 2268\,{\rm s}$.

\subsection{Phase Folded Light Curves}

\begin{figure*}
    \centering
    \includegraphics{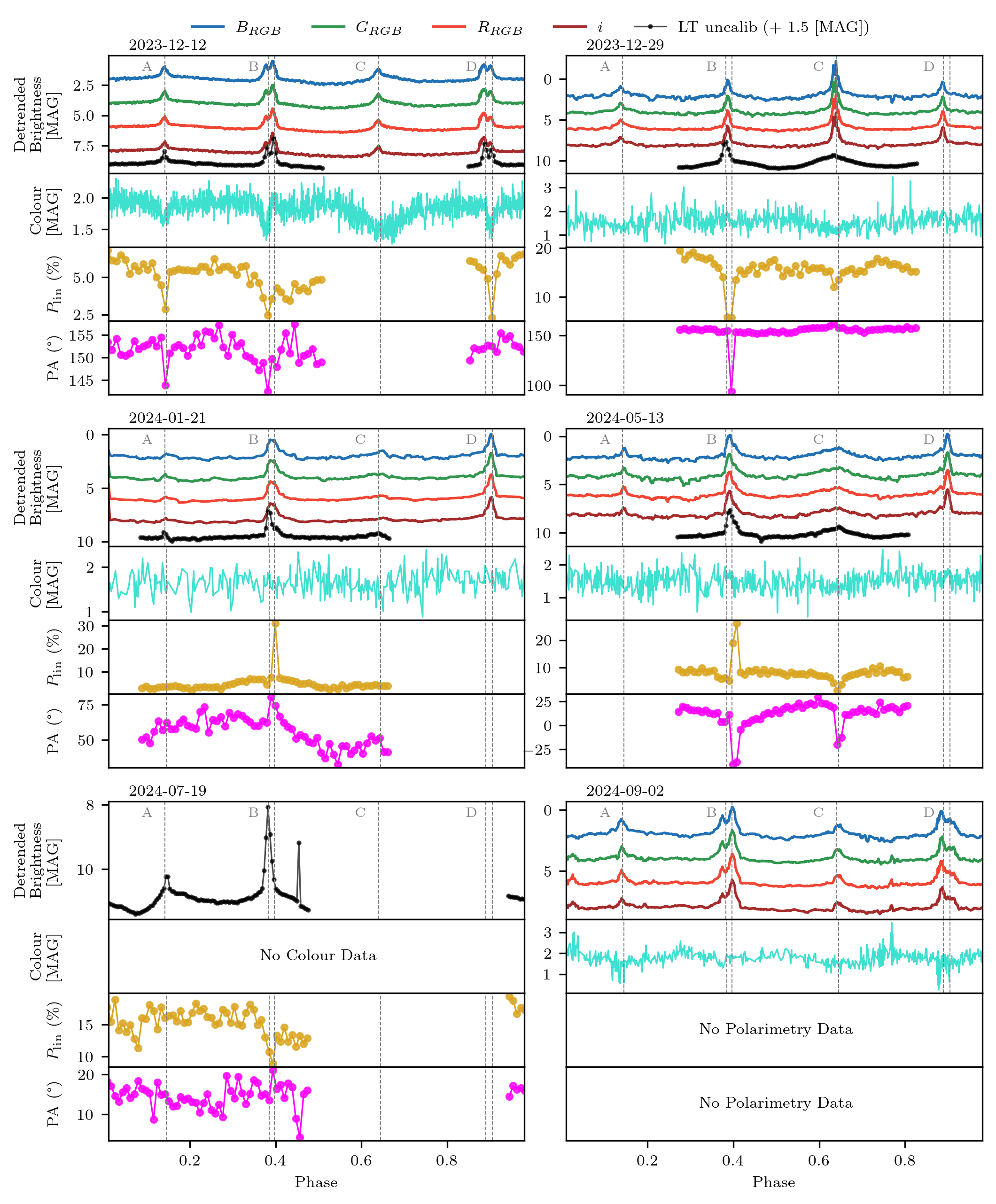}
    \caption{The six observations of Gorizont-6 with the STING telescope and polarimetry measurements (linear polarisation and polarisation angle) from the LT. The photometry (in all four STING bands) and polarimetry measurements are displayed as a function of rotational phase following a phase fold over the estimated bootstrapped rotational periods from the $B_{RGB} - i$ colour. In the case of the unreliable measurement on the night of the 21st January 2024, we take twice the bootstrapped estimate recovered from the $i$ band data. The individual STING band brightnesses were de-trended using the best fit linear baseline coefficients from the analysis prior, re-centred and offset for visual clarity. Like features across the photometry and polarimetry are annotated by the grey dashed vertical lines and are assigned A-D.}
    \label{fig:Phot_Pol}
\end{figure*}

Fig. \ref{fig:Phot_Pol} displays the phase-folded photometric and polarimetric measurements from STING and LT for the six observation epochs. 

Considering only the measurements on the night of 12th December 2023 for the moment, the colour index (\(B_{RGB}-i\)) helps us distinguish between reflections originating from different facets on the spacecraft. In Fig.~\ref{fig:Phot_Pol}, peak~A shows a narrow blueward colour shift compared to peak~C, which is broader and overall bluer. We speculate that peaks~A and~C arise from reflections off the back and front of the solar panels, respectively. The remaining doublet signatures (features~B and~D) are consistent with reflections from two surfaces on the spacecraft body with slightly different surface properties: in each doublet, the fainter of the two peaks is strongly blue, and the ordering of the bright and faint components flips between the two doublets, consistent with viewing the same feature from opposite sides as the object rotates.

We also see that in the following observed epochs, the reflection from the front of the solar panels eventually disappears due to the geometry not being favourable for specular reflections. This behaviour may suggest that the spin-axis has rotated away and is now approaching the polarisation angle for the material on the body given the increase in linear polarisation. Furthermore, we see the specular features begin to return in the latter two epochs.

\subsubsection{Polarisation behaviour}

The panels of Fig. \ref{fig:Phot_Pol} show a rich polarimetric phenomenology, in line with expectations of a rotating object. The epochs with better quality data show a clear polarimetric signature associated with the glints, particularly obvious in the two epochs of December 2023 and May 2024. These generally appear as very short-lived dips in linear polarisation and a short-lived change in polarisation angle (N.B: the polarisation angle is defined in the astronomical system where polarisation angle towards North is $0^{\circ}$ and East is $90^{\circ}$). These  changes within glints are very short-lived, often limited to just one or two polarimetric datapoints. This indicates that the obtained cadence of MOPTOP in {\tt SLOW} mode (20 seconds as explained above), is generally insufficient to fully resolve the glint in polarisation space (four half-wave positions are needed to determine $Q$ and $U$, limiting time resolution). Observations in {\tt FAST} mode (giving a 2 sec cadence) would have likely resolved this better, but at the cost of a loss of signal-to-noise. 

Glints from a reflecting element on a satellite occur when the satellite's flat, reflective surface is oriented such that the sun's reflection (light from the sun) is directed straight towards an observer. In a single specular reflection of a simple material of a satellite we may expect the maximum linear polarisation to occur close to the Brewster angle (e.g. \citealt{Stryjewski}), and the resulting light curve and polarisation curve of an active attitude-controlled satellite may look fairly smooth and have a simple shape with a low polarisation at small phase angles and high polarisation at high phase angles. This has indeed been observed for several satellites (e.g. \citealt{Kosaka}); more detailed, high cadence observations, show deviations from this general behaviour, which often can be ascribed to the contribution of multiple reflecting elements to the received polarisation (Wiersema et al.~in prep.).

For rotating satellites one can use the assumption of single reflection to generate wavelength-dependent  linear polarisation expectations for a variety of shapes and refractive indices, see for example the work of \cite{Stryjewski} who provide examples of a rotating rocket body and the Hubble space telescope. As can be seen in the simulations by \cite{Stryjewski}, the observed polarisation behaviour can be strongly wavelength dependent and dependent on the rotation axis and viewing geometry (see Figures 16 and 18 in  \citealt{Stryjewski}). While the different reflecting elements each have their own observed polarisation behaviour (as they each have their own orientation and material), they are tied together in that the configuration of the satellite is fixed.  The received polarisation of a rotating object  also depends somewhat on the solar phase angle (\citealt{Stryjewski}). For relatively slowly rotating objects like Gorizont-6, the change in solar phase angle over the observed rotation period (or observation period) can be significant enough to play a role.

The polarimetry dataset covers a period of over 6 months, and a wide solar phase angle range. The six observations broadly appear to show a behaviour where the polarisation decreases with decreasing absolute phase angle. This is commonly observed in attitude controlled active satellites (e.g. \citealt{Kosaka,papadogiannakis}); the polarisation curve can be asymmetric with respect to the sign of the phase angle (see e.g. \citealt{papadogiannakis} and Wiersema et al. in prep.). To our knowledge, this behaviour has not been observed in a tumbling satellite before, and may imply a specific component of rotation.  

Each polarimetric observation in Fig. \ref{fig:Phot_Pol} shows strong, distinct features occurring on timescales of minutes. In most cases, changes in polarisation are accompanied by changes in polarisation angle, as expected from variability from reflecting components rotating in and out of view (i.e. they are seen under rapidly changing phase angles).

The two observations of December 2023 and the observation in May 2024 show clear sharp decreases of linear polarisation, occurring on short timescales, consisting of a shallow broad component and a sharp short duration spike, the latter seen in only one or two datapoints. These sharp depolarisation spikes seem to occur close in time to light curve flares (see Fig. \ref{fig:Phot_Pol}), both the broad light curve spike (see the polarisation data of 13th May 2024) as well as the narrow one (see the polarisation data of 29th December 2023). This is consistent with expectation of a light curve glint at an angle that is far from the Brewster angle, contributing mainly unpolarised (or minimally polarized) light, depolarising the received signal. The polarisation signal appears to broadly mimic the general light curve shape of the flare in some cases (e.g. 29th December 2023), though this appears not always the case. The two December 2023 observations also show clear polarisation angle changes associated with the sharp polarisation dips, although the polarisation degree remains sufficiently high that the polarisation angle is still well-defined, consistent with this scenario. It is also clear from the data that the cadence of the polarimetry is too low to resolve these glints in polarisation angle space, and some of these spikes may be missed because of this relatively coarse (20 sec) sampling (19th July 2024). The high ``baseline'' polarisation at large phase angles improves the constraint on the polarisation position angle: in the high-S/N regime,
$\sigma_{\theta} \approx \tfrac{1}{2}\,\sigma_{P}/P_{\rm lin}$ (radians),
whereas for $P_{\rm lin}$ comparable to the $q,u$ noise level the position angle becomes effectively undefined (polarimetric sensitivity limit).

\subsection{Spin Axis Constraint}
\label{sec:spin-state-constraint}
\begin{figure*}[ht!]
    \centering
    \includegraphics{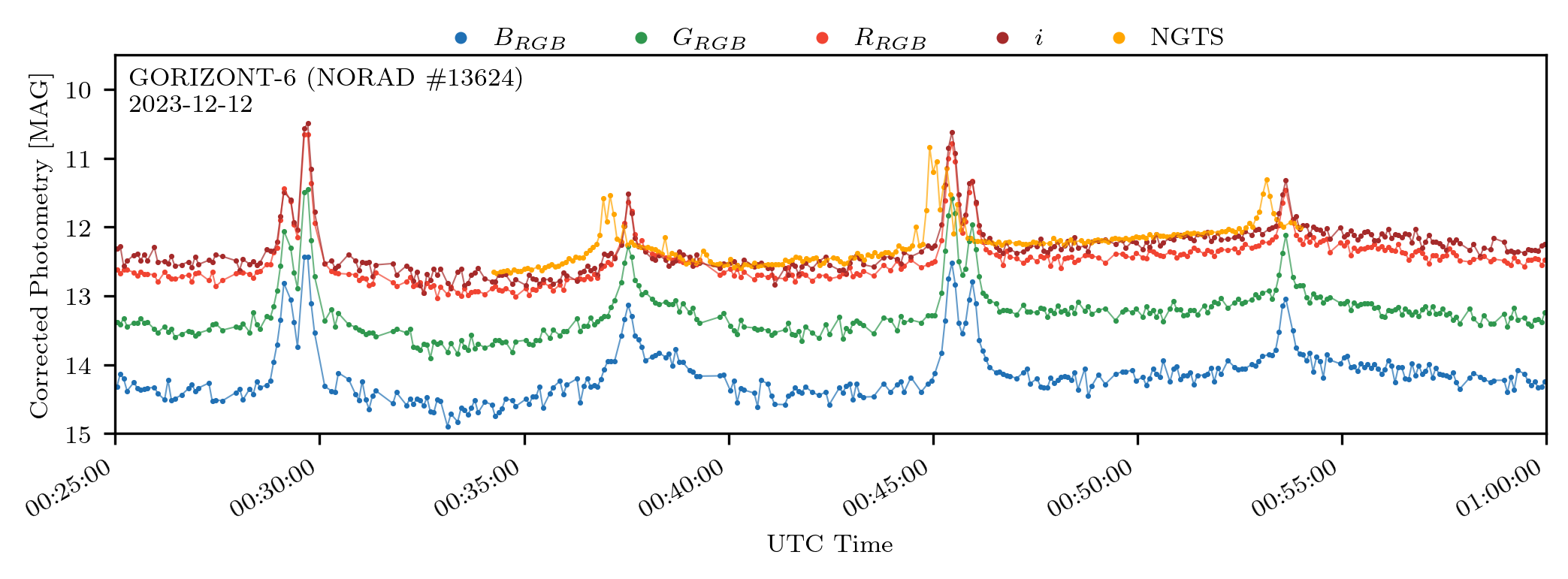}
    \caption{Multi-colour light curves as a function of UTC time. The simultaneous multi-instrument observations of Gorizont-6 were taken on the 12th of December 2023 with STING and NGTS.}
    \label{fig:Multi-obs}
\end{figure*}
Fig. \ref{fig:Multi-obs} displays the light curves from the joint NGTS and STING observation of Gorizont-6 on the night of the 12th December 2023. The glints are visibly offset between the two sites. This offset provides a constraint on the spin-axis, following the methodology and geometrical logic outlined in \cite{2024AdSpR..74.5725K}.

The phase angle bisector (PAB), $p$, is defined as the unit vector halfway between the satellite-to-Sun ($V_{\text{sat$\rightarrow$ Sun}}$) and satellite-to-observer directions ($V_{\text{sat$\rightarrow$obs}}$):
\begin{equation}
    p = \frac{ V_{\text{sat$\rightarrow$obs}}(t) + V_{\text{sat$\rightarrow$ Sun}}(t)}{\| V_{\text{sat$\rightarrow$obs}}(t) + V_{\text{sat$\rightarrow$ Sun}}(t)\|}
    \label{eq:pab_calc}
\end{equation}
It is a useful quantity in determining whether a reflection has occurred off a surface as the specular condition holds true that a reflection will occur when the PAB vector and surface normal, $\vec{n}$ to a flat facet nearly coincide \citep{2013amos.confE..34H}.

\begin{figure*}[ht!]
    \centering
    \includegraphics[
    ]{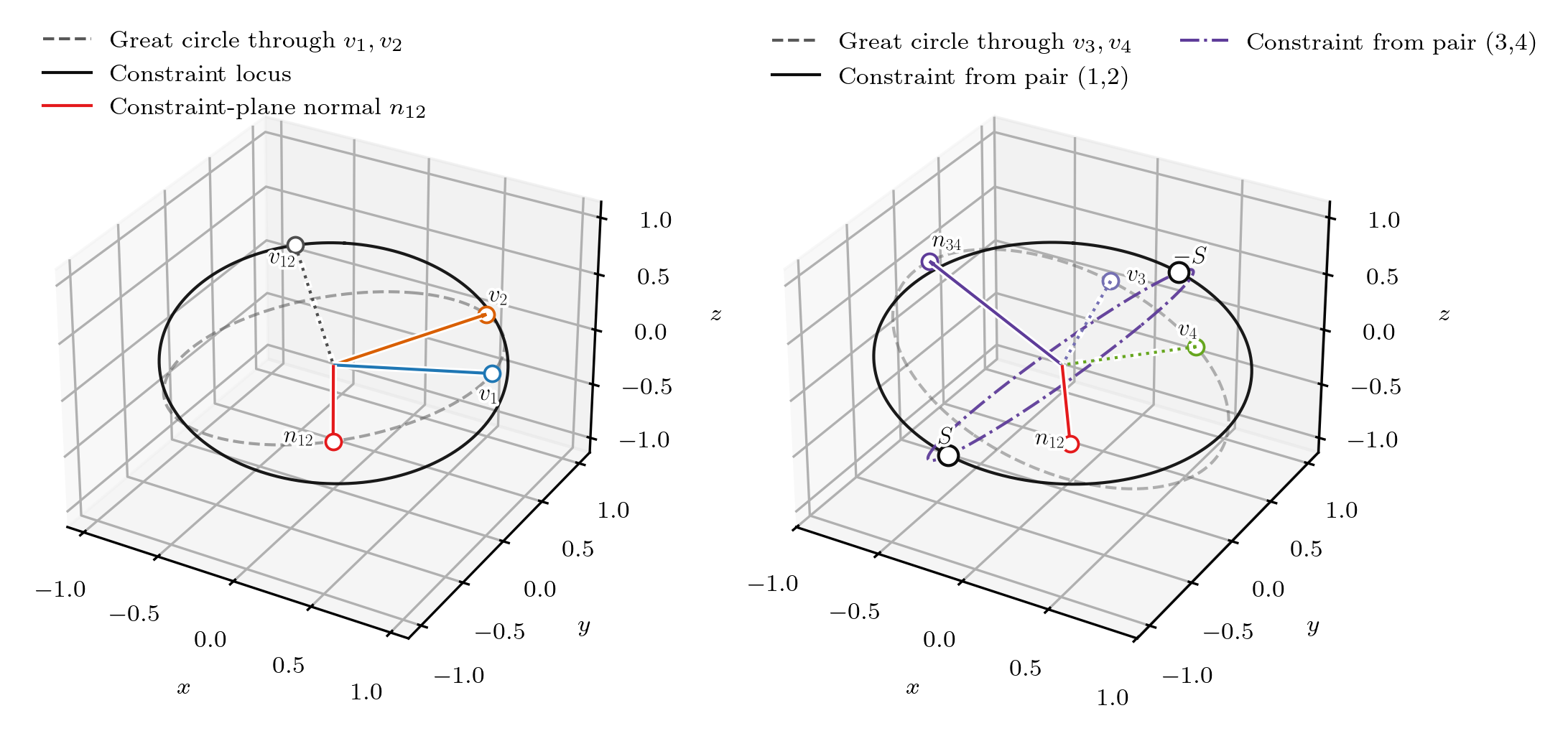}
    \caption{A diagram explaining the spin-pole estimation from matched glint pairs on the unit sphere.
    Unit PAB directions from two observing sites corresponding to the same glint are plotted as vectors $v_i$.
    \textbf{Left:} A matched pair $(v_1,v_2)$ defines the plane spanned by the two directions (dashed great circle) and yields a great-circle locus of admissible spin poles given by the constraint plane with normal
    $n_{12}=b_{12}\times (v_1\times v_2)$, where $b_{12}=v_1+v_2$ is the bisector direction.
    \textbf{Right:} A second matched pair $(v_3,v_4)$ produces a second constraint great circle (purple); the candidate spin axes are the antipodal intersection points $\pm S$, with $S=n_{12}\times n_{34}$.
    A refined estimate is obtained by minimising the cost function in Eq.~\ref{eq:cost} over trial spin-axis directions.}
    \label{fig:spin_axis_glint_diagram}
\end{figure*}

The geometric construction used to obtain the candidate spin-axis directions $\pm S$ from two matched glint pairs is illustrated in Fig.~\ref{fig:spin_axis_glint_diagram}.

The approach is to find the spin-pole as the intersection of two great circles on a unit sphere. Starting with at least two recognised glint events from our two light curves that we can match. We will have PAB vectors from both sites that correspond to that glint, call them $v_1$ and $v_2$, we then calculate the normal to this glint-plane, which is $v_{12} = v_1 \times v_2$ and the bisector vector to the arc that passes through them, $b_{12} =  v_1 + v_2$. We then find the normal vector to these two planes by $n_{12} = b_{12} \times v_{12}$ and we repeat this for another set of glint events occurring in both light curves ending up with a normal vector $n_{34}$. The spin-axis will lie on the line of intersection of these planes, $S = n_{12} \times n_{34}$.

We can then define a cost-function to find a spin-axis that lies at the same pairwise angular distance from the PAB vectors at the instants of the glint events. This cost-function is shown in equation \ref{eq:cost}.
\begin{equation}
    F = \sum(\beta_{i,j}(\alpha,\delta) -\beta_{i}(\alpha,\delta) )^2
    \label{eq:cost}
\end{equation}

\begin{figure}[ht!]
    \centering
    \includegraphics{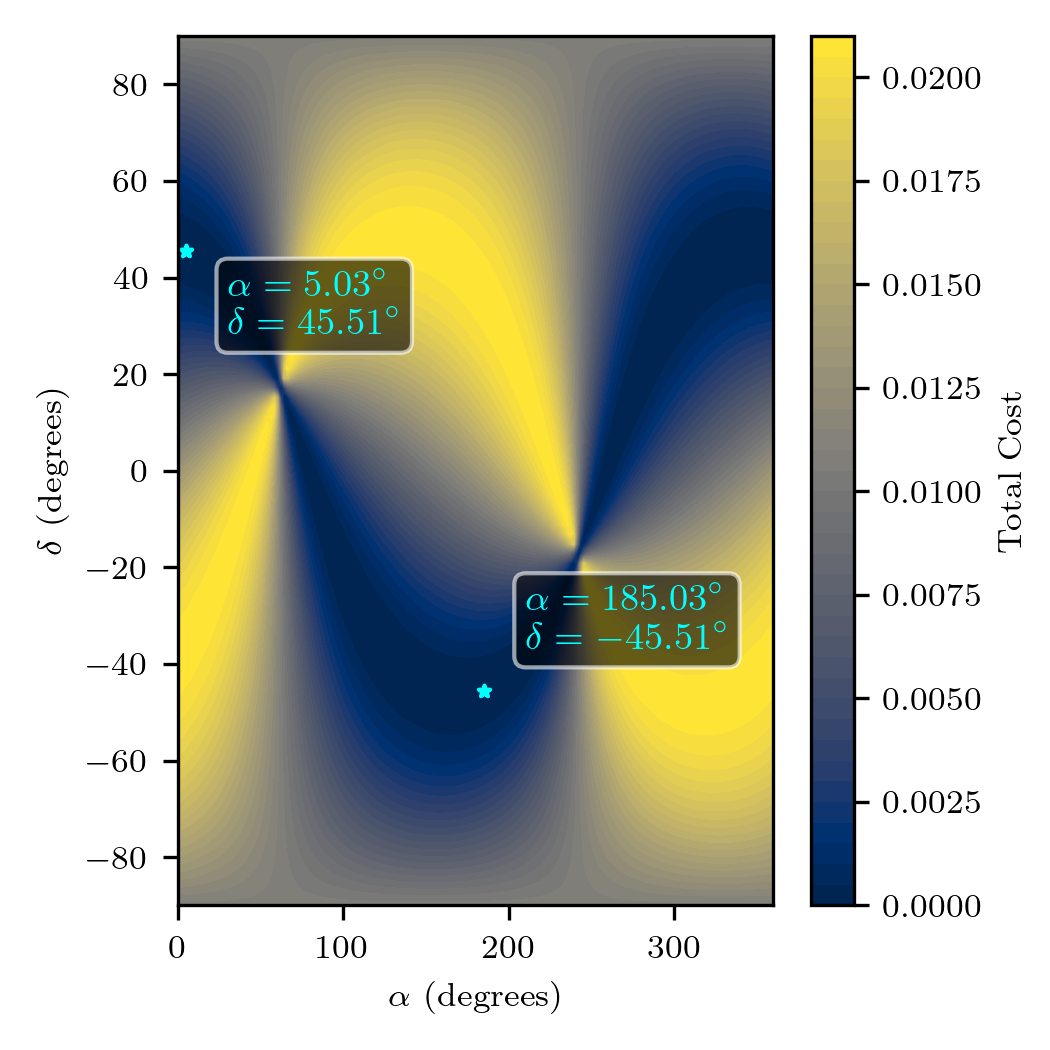}
    \caption{Total cost surface as a function of trial spin-axes constructed from $\alpha$,$\delta$ of the spin-pole in the ICRS. Cyan stars mark the point on the grid where the cost function is minimised and the inversion of this optimised spin-axis. Darker regions are representative of trial spin-axes that are more likely to have caused the observed paired glint events from both sites.}
    \label{fig:Cost_surface_G6}
\end{figure}

Here $\beta_{i,j}$ refers to the angular separation between a trial spin-axis in the ICRS (International Celestial Reference System) and the calculated PAB vector at time of pattern glint. $\beta_i$ refers to the mean PAB latitude of the glint pattern. Given our pattern consists of a glint seen at NGTS and then repeated later by STING, we can only take our mean estimate from these two observations. The resultant cost surface from this least-squares fit is displayed in Fig. \ref{fig:Cost_surface_G6}.

The best fit spin-axis (minimisation of the cost function) occurs at $\alpha = 185.03^\circ$, $\delta = -45.5^\circ$. The antipode of this spin-axis is found at $\alpha = 5.03^\circ$, $\delta = 45.5^\circ$.

\section{Discussion}

The observed spin-period evolution of Gorizont-6 is consistent with the rotational behaviour expected for defunct satellites under the influence of YORP. Unlike asteroids, for which YORP-driven period changes typically occur on Myr timescales \citep[see e.g][]{2021JGRE..12606863K}, inactive satellites can experience appreciable radiative torques on much shorter timescales because of their asymmetric shapes, comparatively small moments of inertia (scales as $MR^2$) relative to illuminated area, and strong contrasts in surface optical properties, such as those between solar panels and the spacecraft bus. The works of \cite{2015AdSpR..56..237A}, \cite{2018AdSpR..61..122A} and \cite{2023AeMiS.102...29C} have extensively simulated the YORP effect for realistic satellite masses and shapes and showed that we can see spin-period changes over relatively short timescales with respect to that seen in asteroids. Over the course of our observations, Gorizont-6 most likely remained in a phase of near-uniform rotation, consistent with the damping trend in its spin period and the morphology of the phase-folded light curves. Previous work has shown that such objects may subsequently return to tumbling states and later re-stabilize \citep{2018AdSpR..61..122A,BensonThesis,Benson2019YORP}. Our results therefore support the interpretation that Gorizont-6 may undergo repeated transitions between stable spin, tumbling, and re-stabilization, as predicted in YORP-driven rotational cycles for defunct satellites.

Our attempt to constrain the spin-pole of Gorizont-6 yielded a range of possible solutions. The minimisation of the least-squares function identifies a plausible spin-pole orientation, but the constraint remains limited by the relatively short observation arcs and the small number of clear glint pairings between the two sites. These limitations would be alleviated by longer observational baselines or by including measurements from a third location.

While wide-field observations from instruments such as STING and NGTS enable the simultaneous monitoring of many objects and provide high precision light curves, they remain fundamentally limited by the inherent degeneracies in photometric inversion. Symmetric spacecraft geometries, harmonic ambiguities in period determination, and the difficulty of associating specific light-curve features with distinct reflecting surfaces all mean that even high-cadence multi-colour photometry cannot always yield a unique rotational solution. In contrast, single-target polarimetry offers a substantially more diagnostic observable: the linear polarisation degree and angle are not subject to the same degeneracies as brightness, and they respond sensitively to changes in viewing geometry, surface specularity, and material refractive indices. Our results for Gorizont-6 demonstrate this clearly—distinct dips in polarisation and rapid swings in polarisation angle pinpoint specific surface interactions even where photometry alone might admit multiple interpretations. Thus polarimetric observations provide the decisive constraints needed to break rotation-state degeneracies, validate the fundamental period, and link light curve features to physical spacecraft components. The optimal strategy for rotational characterisation therefore lies in a hybrid approach: wide-field multi-colour photometry for coverage and cadence, complemented by targeted polarimetry for unambiguous physical interpretation. In this case, we are able to distinguish between components on the main bus and solar panels.

Taken together, our results reinforce the view that the YORP effect does play a role in the long-term rotational dynamics of defunct satellites in graveyard orbits, producing behaviours that parallel those seen in small-body populations. While Gorizont-6 provides a detailed case study, the broader Gorizont series offers an opportunity to test whether satellites sharing a common bus architecture evolve along similar dynamical pathways. In total, 33 Gorizont satellites were launched \citep{krebs_gorizont_1_33}, all built on the KAUR-3 platform; if their large-scale geometries and reflective surfaces are broadly comparable, one would expect recurring light-curve morphologies and preferred spin-state outcomes across the population. To fully evaluate this hypothesis, longer observational baselines, and systematic monitoring of the full series are essential. Looking forward, because spin state strongly shapes light-curve morphology—dictating both periodicity and modulation patterns—multi-colour time-series data from coordinated campaigns could also be leveraged by machine-learning approaches such as boosted decision trees \citep{2024RASTI...3..247S}.

\section{Summary}
We have conducted a coordinated multi-instrument observation campaign combining wide-field photometry with simultaneous high-cadence linear polarimetry to characterise the rotational evolution of the inactive geostationary satellite Gorizont-6. Multi-colour photometry enabled robust period estimation across six epochs, revealing a trend which was best fitted by a damped exponential curve, showing spin-down over nine months, consistent with YORP-driven rotational evolution. The simultaneous polarimetric data provided additional, independent constraints: rapid polarisation dips coincident with photometric glints traced reflections from specific spacecraft surfaces, breaking degeneracies between possible harmonics of the period and allowing us to confidently identify the true rotational period. Combined STING–NGTS observations further enabled a geometric inversion of paired glint events, yielding a plausible range of spin-axis orientations. Future work should involve combining these measurements with BRDF-derived light-curve models to further constrain plausible spin states, alongside dedicated long-term targeted monitoring of Gorizont/KAUR-3-class satellites to probe their dynamical evolution and quantify the impact of YORP on the Graveyard environment.

Taken together, these results demonstrate that simultaneous multi-colour photometry offers a cost-effective and efficient means of obtaining robust period measurements, especially across large satellite samples. However, achieving unambiguous confidence in the true rotation state — and in linking light-curve features to physical components on the spacecraft — ultimately requires polarimetry, the sensitivity of which to surface geometry and material properties provides a unique diagnostic power. Future large-scale surveys of defunct spacecraft will therefore benefit most from a strategy in which wide-field photometry is combined with targeted polarimetry to provide accurate attitude characterisation.

\section*{Acknowledgements}
This work makes use of data from the STING instrument operated on the island of La Palma by the University of Warwick in the Spanish Observatory del Roque de los Muchachos of the Instituto de Astrofísica de Canarias. 
The Liverpool Telescope is operated on the island of La Palma by
Liverpool John Moores University in the Spanish Observatorio del Roque de
los Muchachos of the Instituto de Astrofisica de Canarias with financial support from the UK Science and Technology Facilities Council (STFC). 
This work is based in part on data collected under the NGTS project at the ESO La Silla Paranal Observatory. The NGTS facility is operated by a consortium of institutes with support from the UK Science and Technology Facilities Council (STFC) under projects ST/M001962/1, ST/S002642/1 and ST/W003163/1.
JAB acknowledges support from the Science and Technology Facilities Council
(grant ST/Y50998X/1).

We thank Conor Benson, Antonella Albuja, and Daniel Scheeres (University of Colorado Boulder) for helpful discussions and feedback. KW acknowledges useful discussion with Witold Niemczak and Ella Jefferies. 

For the purpose of open access, the author has applied a Creative Commons Attribution (CC-BY) licence to any Author Accepted Manuscript version arising from this submission. 

The authors would like to thank the reviewers for their comments and suggestions that helped to improve the quality of the final manuscript.

\section*{Data availability}
The data underlying this article will be shared on reasonable request to \texttt{P.Chote@warwick.ac.uk}.
%% Bibliography
%% Author year style
\bibliographystyle{jasr-model5-names}
\biboptions{authoryear}
\bibliography{refs}

\end{document}